*Article*

# Estimating Blood Pressure from Photoplethysmogram Signal and Demographic Features using Machine Learning Techniques

**[1]Moajjem Hossain Chowdhury, [1]Md Nazmul Islam Shuzan, [2]Muhammad E.H. Chowdhury\*, [3]Zaid B Mahbub, [3]M. Monir Uddin, [2,4]Amith Khandakar, [4]Mamun Bin Ibne Reaz**

[1]Department of Electrical and Computer Engineering, North South University, Dhaka 1229, Bangladesh
moajjem.hossain@northsouth.edu, nazmul.shuzan@northsouth.edu

[2]Department of Electrical Engineering, Qatar University, Doha 2713, Qatar
mchowdhury@qu.edu.qa, amitk@qu.edu.qa

[3]Department of Mathematics and Physics, North South University, Dhaka 1229, Bangladesh
zaid.mahbub@northsouth.edu, monir.uddin@northsouth.edu

[4]Department of Electrical, Electronic & Systems Engineering, Universiti Kebangsaan Malaysia, Bangi Selangor 43600, Malaysia mamun@ukm.edu.my

\*Correspondence: Muhammad E.H. Chowdhury (M.E.H.C.), mchowdhury@qu.edu.qa; Tel.: +974-31010775

**Abstract:** Hypertension is a potentially unsafe health ailment, which can be indicated directly from the Blood pressure (BP). Hypertension always leads to other health complications. Continuous monitoring of BP is very important; however, cuff-based BP measurements are discrete and uncomfortable to the user. To address this need, a cuff-less, continuous and a non-invasive BP measurement system is proposed using Photoplethysmogram (PPG) signal and demographic features using machine learning (ML) algorithms. PPG signals were acquired from 219 subjects, which undergo pre-processing and feature extraction steps. Time, frequency and time-frequency domain features were extracted from the PPG and their derivative signals. Feature selection techniques were used to reduce the computational complexity and to decrease the chance of over-fitting the ML algorithms. The features were then used to train and evaluate ML algorithms. The best regression models were selected for Systolic BP (SBP) and Diastolic BP (DBP) estimation individually. Gaussian Process Regression (GPR) along with ReliefF feature selection algorithm outperforms other algorithms in estimating SBP and DBP with a root-mean-square error (RMSE) of 6.74 and 3.59 respectively. This ML model can be implemented in hardware systems to continuously monitor BP and avoid any critical health conditions due to sudden changes.

## 1. Introduction

Measuring blood pressure (BP) is an important aspect in monitoring the health of a person. High blood pressure, generally, means that a person has a higher risk of health problems [1]. High blood pressure puts a huge amount of strain on the arteries and the heart. This strain can make the arteries less flexible over time. As they become more inflexible, the lumen become narrower. Therefore, the probability of it being clogged up (clot) increases. A clot is very dangerous and may cause heart attack, stroke, kidney diseases and dementia. As a result, it is important for a person to monitor their blood pressure regularly. In most cases, measuring blood pressure once or twice a day is more than enough.

However, sometimes the doctor needs to track the blood pressure continuously. This is because blood pressure is known to decrease at night. So, it is useful to measure the blood pressure overnight, as an abnormal dip in blood pressure may suggest a higher risk of cardiovascular problems [2].

The current standard methods include either a cuff based BP measurement or an invasive procedure for BP measurement. The cuff method measures the blood pressure after a set interval (e.g., of 15 minutes). This means that the end-result is discrete and uncomfortable to the user. Furthermore, this process requires the arm to be kept steady while the inflation and deflation causes disturbance in patient's sleep. Arterial lines management is an invasive procedure that allow for continuous blood pressure monitoring. However, the invasive procedure leaves the patient vulnerable to infection. Hence, there is a need for a non-invasive, cuff-less, continuous BP monitoring system. With the advent of digital sensors, signal-processing, machine-learning algorithms and advanced physiological models help to gather important human vital signs using wearable sensors [3, 4]. Even the indirect estimation of blood pressure (BP) using Photoplethysmography (PPG) has become more realistic [5-8].

Photoplethysmography (PPG) was being used for decades for measuring the amount of light absorbed or reflected by blood vessels in the living tissue. PPG technology is a versatile and low-cost technology [9], which can be extended to different aspects of cardiovascular surveillance including identification of blood oxygen saturation, heart rate, BP estimation, cardiac output, respiration, arterial ageing, endothelial control, micro-vascular blood flow, and autonomic function [10]. Many different kinds of PPG signals have been identified and have been shown associated with age and cardiovascular pathology [11, 12]. In clinical practice, PPG signals are recorded from micro-vascular beds at exterior body locations, such as the finger, earlobe, forehead, and toe [13]. The coverage area of the PPG sensor includes veins, arteries and numerous capillaries. PPG waveforms generally have three distinct features. As shown in **Figure 1**, a PPG waveform typically contains systolic peak, diastolic peak and a notch in between.

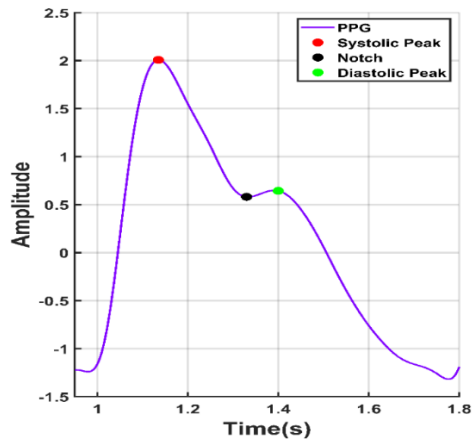

**Figure 1:** A typical PPG waveform with notch, systolic peak and diastolic peak.

The raw PPG signal typically includes pulsatile and non-pulsatile blood volumes [14]. The pulsatile portion of the PPG signal is attributed to the variation in blood pressure within the arteries and is synchronous to the pulse, while the non-pulsating part is a result of normal blood volume, respiration, sympathetic nervous system, and thermoregulation [15]. Green, red and infrared light are often used to extract PPG waveforms. Red and infrared light can reach approximately 2.5 mm, whereas green light can penetrate less than 1 mm into the tissue [16]. Therefore, infrared light is typically used for acquiring PPG signal for the measurement of blood pressure. Although the PPG tool is a low-cost and portable optical electronic device, its measurement has several challenges, such as, noise reduction [17-19] and multi-photodetector creation [20].

Several techniques to estimate BP from PPG were proposed in the recent works. Some algorithms [21] incorporate waveform analysis and biometrics of PPG to estimate BP, which has been tested in subjects with different age, height and weight. When calibrated, PPG shows great potential to track BP fluctuations, which can bring enormous health and economic benefit. An easy and bio-inspired mathematical model was proposed at [22] to predict estimating Systolic BP (SBP) and Diastolic BP (DBP) through careful mathematical analysis of the PPG signals. Systolic and diastolic blood pressure levels were predicted using Pulse Transit Time (PTT) in [23, 24] and combination of Paroxysmal Atrial Tachycardia (PAT) and heart rate in [25], while combination showed improvement over PTT alone. The beat-to-beat optical BP measurement method was developed, tested and reported using only PPG from fingertips [26]. Key features such as amplitudes and cardiac part phases were extracted through a fast Fourier transformation (FFT) and used to train an artificial neural network (ANN), which was then used to estimate BP using PPG. In [27], support vector machine (SVM) algorithm showed better accuracy than the linear regression method and ANN.

The recent growth in the field of deep learning has made it potential for this application. Su et al. 2018 [28] discussed the problem of accuracy reduction in the current models for BP estimation from PPG due to the requirement of frequent calibration. A deep recurrent neural network (RNN) with long short-term memory (LSTM) was used to create a model for the time-series BP data. PPG and Electrocardiogram (ECG) were taken as inputs, and PTT with some other features were used as predictors to estimate BP. This method showed improvements in BP prediction compared to other existing methods. Gotlibovych et al. investigated the potential of using raw PPG data to detect arrhythmia in 2018 [29] with reasonable success, which shows the possibility of using raw PPG signal as inputs to the deep learners. In [30], the authors have created a novel spectro-temporal deep neural network that took the PPG signal and its first and second derivative as inputs. The neural network model had residual connections and were able to get mean absolute error (MAE) of 6.88 and 9.43 for DBP and SBP, respectively.

Several research groups have analyzed and evaluated the quality of the open-source dataset, which was used in this study [18, 30-32]. A novel approach [33] for treating hypertension based on the theory of arterial wave propagation and morphological theory of PPG was proposed to check the physiological changes in different levels of blood pressure. ECG and PPG signals were obtained simultaneously to detect hypertension. A model for PPG characteristic was analyzed and an inherent relationship between the characteristics of Systolic BP and PPG was established [34]. In [35], PPG signal analysis was used to characterize obesity, age group and hypertension using PPG pulse based on the pulse decomposition analysis.

The features typically used for non-invasively estimating BP are: (i) t-domain, (ii) f-domain, (iii) (t, f)-domain, (iv) and statistical features. Several t-domain features, which were calculated from the original signal and its derivatives, were used by different groups [9, 36-38]. In a different study, Zaid et al. [39] showed the use of frequency domain features for identifying neurological disorder and in this study, the authors have taken inspiration from Zaid et al. to create features in estimating BP accurately from the PPG signal.

Several studies reported different features of PPG signal for different application [9, 34, 38, 40]. Various groups have used these features for SBP and DBP measurement; however, there is still plenty of scope for improvement. Numerous automated ML techniques were evaluated and recorded for various PPG databases as mentioned earlier. Nonetheless, to the best of our knowledge, no recent work has combined t-, f- and (t, f) domain features to estimate BP with high accuracy using machine-learning approach. PPG signal processing is comparatively simpler and easier, so more attention is being paid to novel methods that extract features from PPG signals. To reduce the error in BP estimation based on the PPG signal, this analysis not only extracts features from the PPG signal but also utilizes the demographic characteristics of subjects, such as height, weight and age etc. There are several features were extracted for BP estimation from PPG signal in this study, which were not used before by any other group.

The manuscript is divided into four sections. **Section 1** is discussing the basics of the PPG signal, related works and inspirations of this research. Methodology and database are presented in **Section 2** along with pre-processing steps and system assessment. **Section 3** summarizes the result and discusses the results while **Section 4** concludes the work.

## 2. Materials and Methods

This section discusses about the dataset used in the study, the signal pre-processing techniques used, the features extracted, feature selection techniques used, and the machine learning algorithms models trained and tested to estimate SBP and DBP.

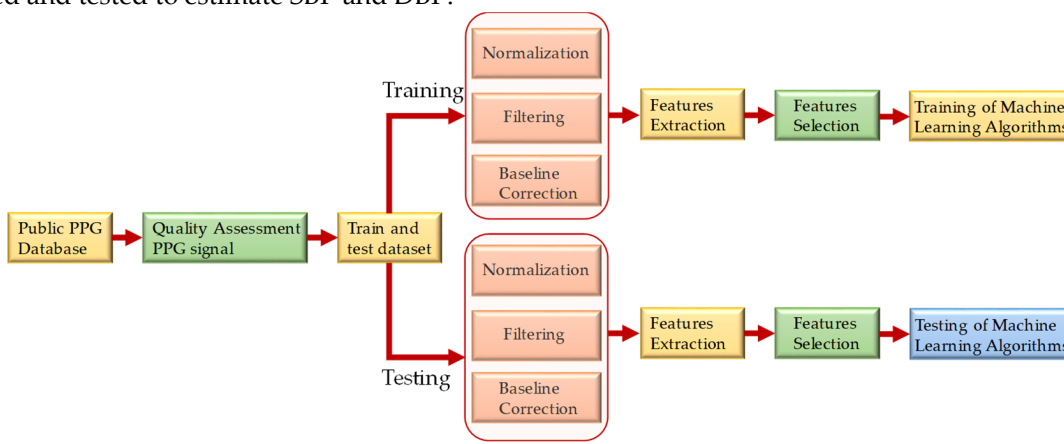

**Figure 2:** Overall system block diagram

*As shown in **Figure 2**, PPG signals were first assessed to check signal quality and then randomly divided into two sets. 85% of the data was used for training and validation and 15% of the data was used for testing the performance of the model. The PPG signals were pre-processed before they were sent for feature extraction. After extracting meaningful features, feature selection techniques were used to reduce computational complexity and chance of over-fitting the algorithm. The features were then used to train machine-learning algorithms. The best regression model was selected for SBP and DBP estimation individually.2.1 Dataset Description*

The dataset used in this study was taken from Liang et al. [31], which is publicly available. The dataset contained 657 PPG signal samples from 219 subjects [18]. The PPG signal were sampled at a rate of 1000Hz and contained 2100 data points per signal with a signal duration of 2.1s. Other than PPG signal, patients' demographic information such as age, gender, height, and weight along with systolic pressure, diastolic pressure, and heart rate were also recorded. Summary of the dataset is shown in **Table 1**.

**Table 1.** Data Summary

| Physical Index | Numerical Data |
|---|---|
| Females | 115 (52%) |
| Age (years) | 57±15 |
| Height (cm) | 161±8 |
| Weight (kg) | 60±11 |
| Body Mass Index (kg/m2) | 23±4 |
| Systolic Blood Pressure (mmHg) | 127±20 |
| Diastolic Blood Pressure (mmHg) | 71±11 |
| Heart Rate (beats/min) | 73±10 |

Of the 657 signals, many signals were of poor quality and could not be used for feature extraction. Liang et al. [18] used a skewness-based Signal Quality Index (SQI) to find the suitable signals. In the quality assurance process, 222 signals from 126 subjects were finally kept for this study. **Figure 3** shows the sample PPG signal which were divided as fit and unfit for the study. It is obvious that the unfit waveforms either do not have prominent features or the diastolic part of the waveform is not obvious in the recorded signal and the data length is very short. Hence, they were not used for the study.

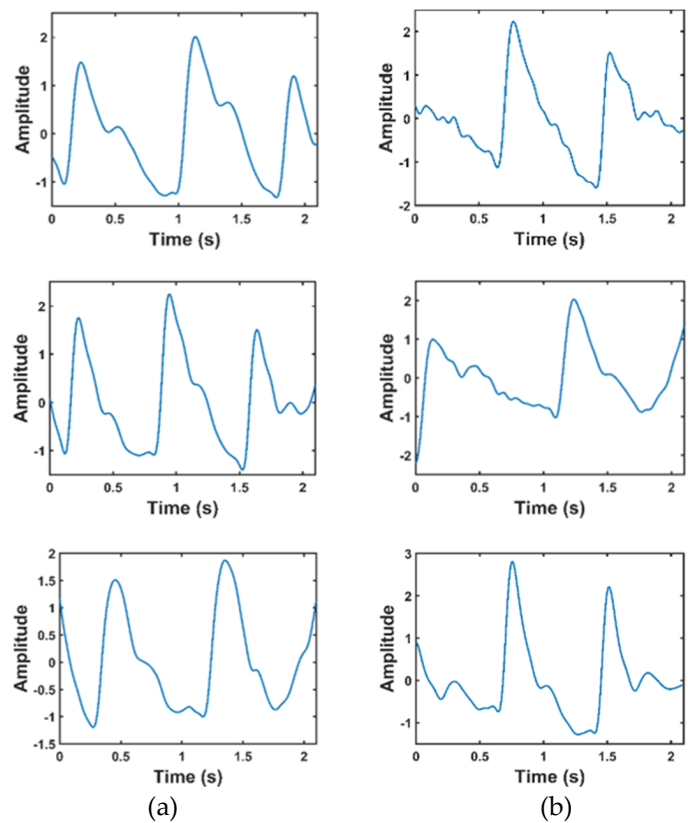

(a)　　　　　　　　(b)

**Figure 3:** Comparison of waveforms that are fit and unfit for study, (a) fit data, (b) unfit data.

*2.2 Preprocessing Signals*

The raw PPG signals were prepared through different pre-processing stages before feature extraction, which are summarized below:

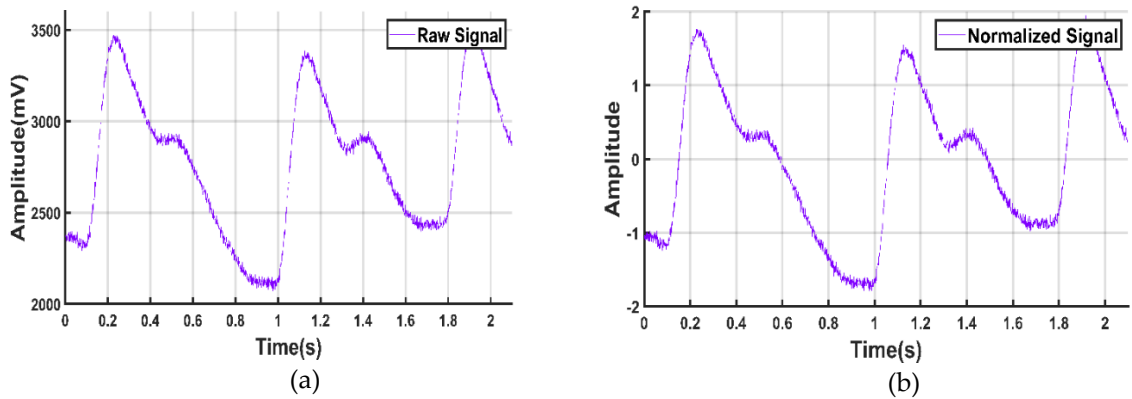

**Figure 4**: PPG Signal: (a) before normalization, (b) after normalization.

### 2.2.1 Normalization

To extract meaningful information from the signals, it was necessary to normalize all the signals. The Z-score technique was used to normalize the signals in this study to get amplitude-limited data.

$$\text{Z-score Normalized Signal} = \frac{Signal - Signal\ Mean}{Standard\ Deviation\ of\ Signal} \qquad (1)$$

It was also observed that after normalization, other pre-processing techniques were easier to implement. **Figure 4** shows the sample PPG signal before and after normalization.

### 2.2.2 Signal Filtration

It was observed that, the signal from the database [31] has high-frequency noise components. Thus, the signals were filtered through low-pass filter that can remove these high-frequency components. Several filtration techniques were tested to de-noise the signal, such as, moving average, low pass finite impulse response (FIR) and Butterworth Infinite Impulse Response (IIR) Zero-Phase Filter. **Figure 5** shows the raw signal overlaid with the filtered output using different type of filters. From the figure 5, we can see that the Butterworth filter produced the filtration. Hence, we used it to filter the PPG waveforms. Which was also used by others to remove noise from the PPG signals [9, 12, 37, 41]. In this work, sixth order IIR filter with a cut-off frequency of 25 Hz was designed in MATLAB.

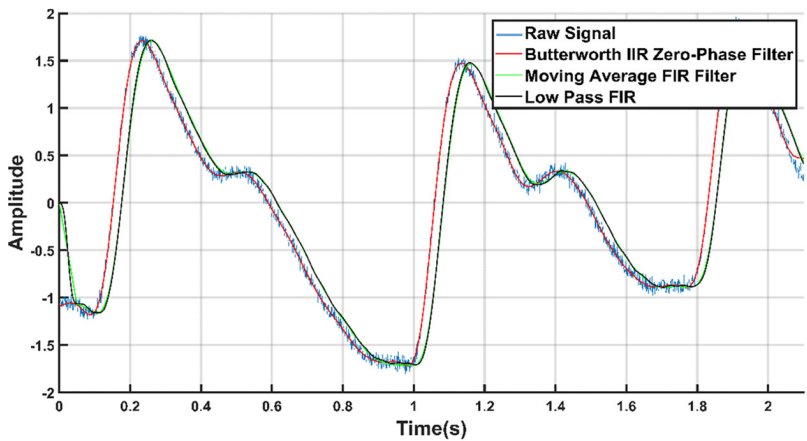

**Figure 5**: Filtered Signals overlaid on the raw PPG signal.

### 2.2.3 Baseline Correction

The PPG waveform is commonly contaminated with baseline wandering due to respiration at frequencies ranging from 0.15 to 0.5 Hz [11, 21, 42, 43]. It is therefore very important that the signal is properly filtered to remove Baseline Wandering but that important information is preserved as far as possible. We used polynomial fit to find the trend in the signal. Then we subtracted the trend to get the baseline corrected signal as shown in Figure 6.

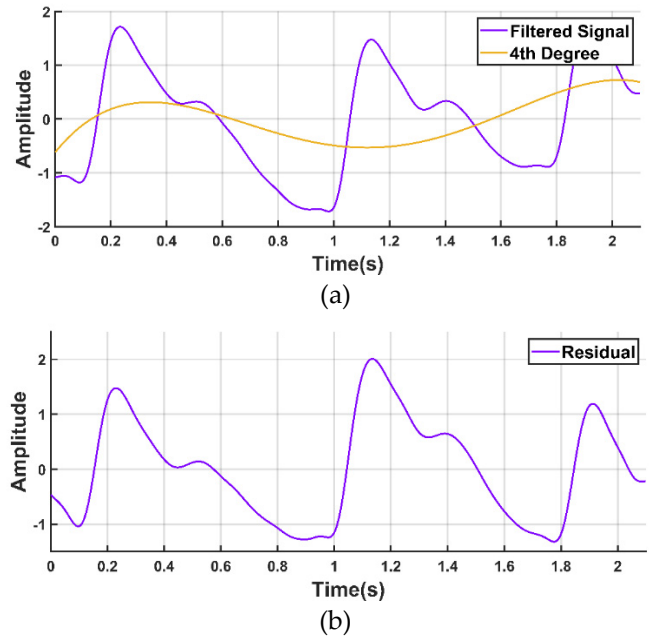

(a)

(b)

**Figure 6**: Baseline Correction of PPG Waveform (a) PPG Waveform with baseline wander and 4th degree polynomial trend, (b) PPG Waveform after de-trending.

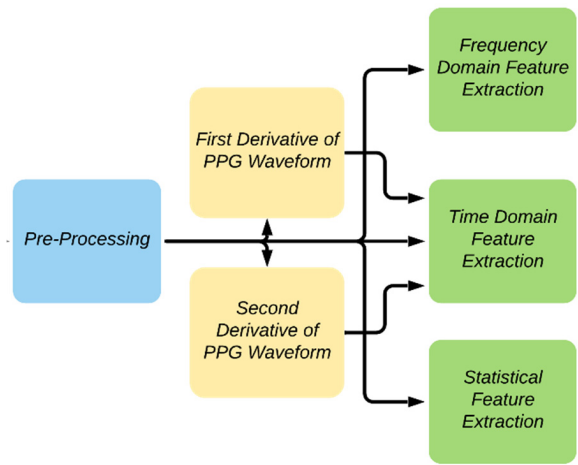

**Figure 7:** Overview of Feature Extraction.

### 2.3 Feature Extraction

The block diagram summarizing the feature extraction details adopted in the study is shown in Figure 7. A PPG waveform contains many informative information such as systole, diastole, notch, pulse width, peak-to-peak interval etc. Some of the distinctive features of PPG waveform might be not dominant in some patients, such as the notch prevalence changing with age[44]. To find the different key points of PPG signal as shown in **Figure 11**(a), the authors have followed the methods described in previous work [45]. The technique was largely based on the derivatives and thresholds defined in [46] and [47].

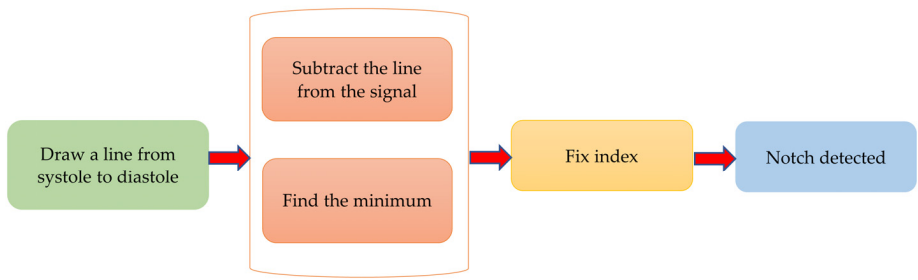

**Figure 8:** Algorithm of notch detection.

The dicrotic notch is an essential feature of the PPG signal. **Figure 8** describes the algorithm to detect the dicrotic notch. To do so, a line was drawn from the systolic peak to the diastolic peak. The minimum of the subtraction of the straight line from the signal is the dicrotic notch. However, to make it more robust, Fix index was used which calculates the local minima within a given window (in this case 50ms) around a given point. Reliable detection of dicrotic notch in various situations is shown in the **Figure 9**.

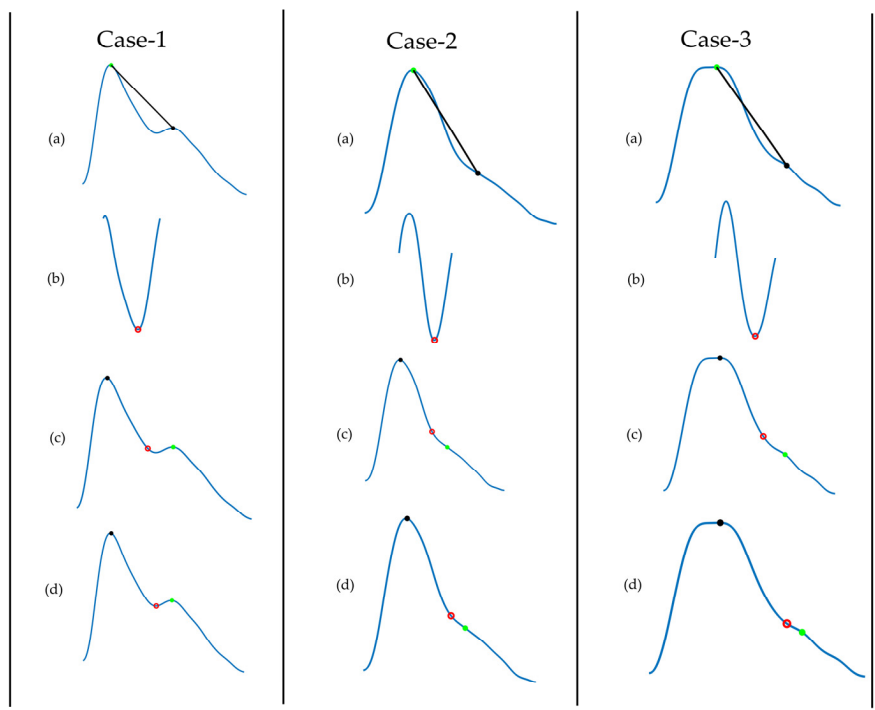

**Figure 9:** Demonstration of dicrotic notch detection for different age group: Case 1 (26 years) ,2 (45 years), and 3 (80 years): (a) Filtered PPG signal where we draw a line from systolic peak to diastolic peak; (b) subtract the line from the signal and find its minimum point; (c) initial notch detected; (d) adjust the notch using fix index.

Another key feature is the foot of the PPG signal. To find the foot of the PPG waveform, the second derivative of the PPG waveform, also called APG (acceleration plethysmogram), was first calculated. From the APG, zone of interest was defined, where the moving average of APG is larger than an adaptive threshold. In the zone of interest, the highest point of the APG corresponds to the foot of the signal. This method is robust and allows detecting the foot of the signal very accurately. **Figure 10** shows that the algorithm can detect prominent foot and flat foot accurately [45].

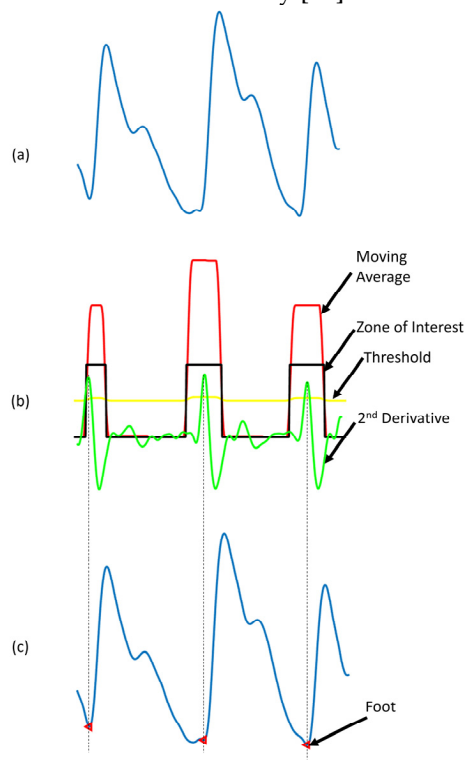

**Figure 10:** Detection of the foot of a PPG waveform; (a) Filtered PPG signal; (b) 2nd derivative of PPG along with derivation of zone of Interest based on moving average of APG and adaptive threshold; (c) Foot of the signal detected.

PPG signal's first and second derivatives were calculated and the relationship between PPG signals and their first and second derivatives is shown in **Figures 11**. The PPG signal is analyzed to extract a1, b1 point from its first derivative and a2, b2 point from second derivative. **Figure 12** shows the frequency domain representation of the PPG signal. Frequency domain representation was analyzed and features related to the first three peaks were extracted. The length of the fast Fourier transform was 2100, which was equal to the number of data points in the signal. Furthermore, demographic data such a Height, Weight, BMI, Gender, Age and Heart Rate were also used as features. It was reported by several groups that demographic features are important features for BP estimation [48]. Elgendi [9] emphasized the need of height details for accurate estimation of PPG waveform while Kavasaoglu et al. [36] found that demographic features were useful and highly ranked features in their Machine Learning Algorithm using PPG signal's characteristics features. In real-time scenario, age and BMI will be known to the user and heart rate can be easily calculated from the PPG signal. Definitions of the extracted time-domain and demographic features were listed in **Tables 2, 3, 4** and **5**. Frequency-domain and statistical features can significantly contribute in BP estimation and were defined in **Tables 6** and **7** respectively. Therefore, 107 features encompassing seventy-five t-domain, sixteen f-domain, and ten statistical features were derived for each PPG signal along with six demographic data. The t-domain, f-domain and statistical features were identified from different previous works [3, 4, 9, 23, 25-27, 38, 39]. It is reported in literature that 1-24 and 42-58 features were used in PPG related works [49]. These features are considered as Literature Features in **Section 3.**

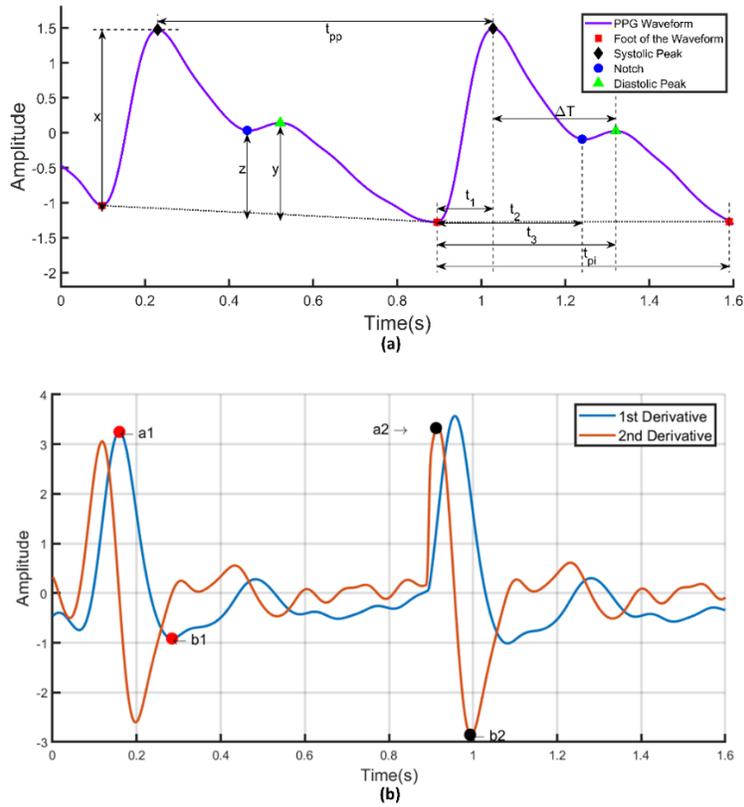

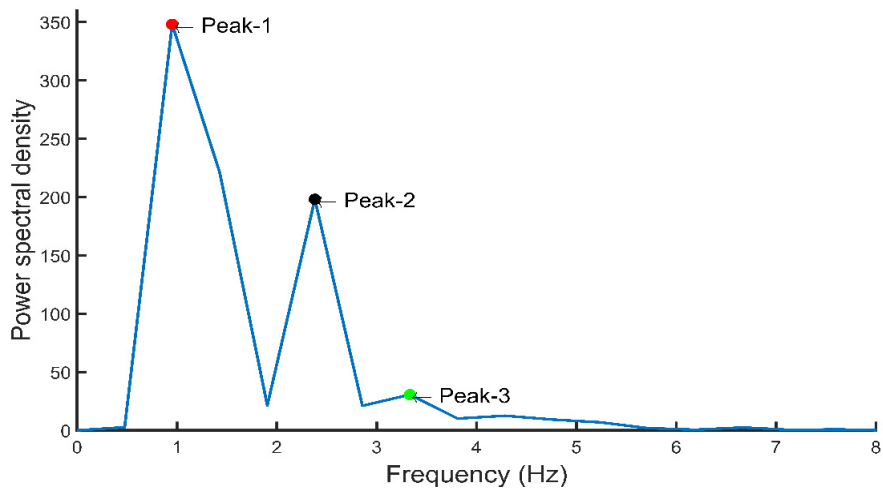

**Figure 11:** (a) Illustration of Time domain features in a PPG signal; (b) First and Second derivatives of PPG signal.

**Figure 12:** Frequency domain representation of PPG signal with important features.

**Table 2.** Twenty-four Features from PPG signal

| Feature | Definition |
|---|---|
| 1. Systolic Peak | The amplitude of ('x') from PPG waveform |
| 2. Diastolic Peak | The amplitude of ('y') from PPG waveform |
| 3. Height of Notch | The amplitude of ('z') from PPG waveform |
| 4. Systolic Peak Time | The time interval from the foot of the waveform to the systolic peak ('$t_1$') |
| 5. Diastolic Peak Time | The time interval from the foot of the waveform to the height of notch ('$t_2$') |
| 6. Height of Notch Time | The time interval from the foot of the waveform to the diastolic peak ('$t_3$') |
| 7. ΔT | The time interval from systolic peak time to diastolic peak time |
| 8. Pulse Interval | The distance between the beginning and the end of the PPG waveform ('$t_{pi}$') |
| 9. Peak to Peak Interval | The distance between two consecutive systolic peaks ($t_{PP}$) |
| 10. Pulse Width | The half-height of the systolic peak |
| 11. Inflection Point Area | The waveform is first split into two parts at the notch point. The area of the first part is $A_1$ and the area of the second part is $A_2$. The ratio of $A_1$ and $A_2$ is the inflection point area ('$A_1/A_2$') |
| 12. Augmentation Index | The ratio of diastolic and systolic peak amplitude ('y/x') |
| 13. Alternative Augmentation Index | The difference between systolic and diastolic peak amplitude divided by systolic peak amplitude ('(x-y)/x') |
| 14. Systolic Peak Output Curve | The ratio of systolic peak time to systolic peak amplitude ('$t_1$/x') |
| 15. Diastolic Peak Downward Curve | The ratio of diastolic peak amplitude to the differences between pulse interval and height of notch time ('y/ $t_{pi}$- $t_3$') |
| 16. $t_1/t_{PP}$ | The ratio of systolic peak time to the peak-to-peak interval of the PPG waveform |
| 17. $t_2/t_{PP}$ | The ratio of notch time to the peak-to-peak interval of the PPG waveform |
| 18. $t_3/t_{PP}$ | The ratio of diastolic peak time to the peak-to-peak interval of the PPG waveform |
| 19. $\Delta T/t_{PP}$ | The ratio of ΔT to the peak-to-peak interval of the PPG waveform |
| 20. z/x | The ratio of the height of notch to the systolic peak amplitude |
| 21. $t_2$/z | The ratio of the notch time to the height of notch |
| 22. $t_3$/y | The ratio of the diastolic peak time to the diastolic peak amplitude |
| 23. x/($t_{pi}$-$t_1$) | The ratio of systolic peak amplitude to the difference between pulse interval and systolic peak time |
| 24. z/($t_{pi}$-$t_2$) | The ratio of the height of notch to the difference between pulse interval and notch time |

**Table 3.** Seventeen Width related PPG Features

| Feature | Definition |
|---|---|
| 25. Width(25%) | The width of the waveform at 25% amplitude of systolic amplitude |
| 26. Width(75%) | The width of the waveform at 75% amplitude of systolic amplitude |
| 27. Width(25%)/t1 | The ratio of pulse width at 25% of systolic amplitude to systolic peak time |

| 28. Width(25%)/t2 | The ratio of pulse width at 25% of systolic amplitude to notch time |
|---|---|
| 29. Width(25%)/t3 | The ratio of pulse width at 25% of systolic amplitude to diastolic peak time |
| 30. Width(25%)/$\Delta$T | The ratio of pulse width at 25% of systolic amplitude to $\Delta$T |
| 31. Width(25%)/tpi | The ratio of pulse width at 25% of systolic amplitude to pulse interval |
| 32. Width(50%)/t1 | The ratio of pulse width at 50% of systolic amplitude to systolic peak time |
| 33. Width(50%)/t2 | The ratio of pulse width at 50% of systolic amplitude to notch time |
| 34. Width(50%)/t3 | The ratio of pulse width at 50% of systolic amplitude to diastolic peak time |
| 35. Width(50%)/$\Delta$T | The ratio of pulse width at 50% of systolic amplitude to $\Delta$T |
| 36. Width(50%)/tpi | The ratio of pulse width at 50% of systolic amplitude to pulse interval |
| 37. Width(75%)/t1 | The ratio of pulse width at 75% of systolic amplitude to systolic peak time |
| 38. Width(75%)/t2 | The ratio of pulse width at 75% of systolic amplitude to notch time |
| 39. Width(75%)/t3 | The ratio of pulse width at 75% of systolic amplitude to diastolic peak time |
| 40. Width(75%)/$\Delta$T | The ratio of pulse width at 75% of systolic amplitude to $\Delta$T |
| 41. Width(75%)/tpi | The ratio of pulse width at 75% of systolic amplitude to pulse interval |

**Table 4.** Sixteen Features derived from first and second derivative

| Feature | Definition |
|---|---|
| 42. $a_1$ | The first maximum peak from the 1st derivative of the PPG waveform |
| 43. $t_{a1}$ | The time interval from the foot of the PPG waveform to the time at which $a_1$ occurred |
| 44. $a_2$ | The first maximum peak from the 2nd derivative of the PPG waveform after $a_1$ |
| 45. $t_{a2}$ | The time interval from the foot of the PPG waveform to the time at which $a_2$ occurred |
| 46. $b_1$ | The first minimum peak from the 1st derivative of the PPG waveform after $a_1$ |
| 47. $t_{b1}$ | The time interval from the foot of the PPG waveform to the time at which $b_1$ occurred |
| 48. $b_2$ | The first minimum peak from the 2nd derivative of the PPG waveform after $a_2$ |
| 49. $t_{b2}$ | The time interval from the foot of the PPG waveform to the time at which $b_2$ occurred |
| 50. $b_2/a_2$ | The ratio of $b_2$ to $a_2$ |
| 51. $b_1/a_1$ | The ratio of First minimum peak of 1st Derivative after $a_1$ to first maximum peak of 1st Derivative |
| 52. $t_{a1}/t_{pp}$ | The ratio of $t_{a1}$ to the peak-to-peak interval of the PPG waveform |
| 53. $t_{b1}/t_{pp}$ | The ratio of $t_{b1}$ to the peak-to-peak interval of the PPG waveform |
| 54. $t_{b2}/t_{pp}$ | The ratio of $t_{b2}$ to the peak-to-peak interval of the PPG waveform |
| 55. $t_{a2}/t_{pp}$ | The ratio of $t_{a2}$ to the peak-to-peak interval of the PPG waveform |
| 56. $(t_{a1} - t_{a2})/t_{pp}$ | The ratio of the difference between $t_{a1}$ and $t_{a2}$ to the peak-to-peak interval of the PPG waveform |
| 57. $(t_{b1} - t_{b2})/t_{pp}$ | The ratio of the difference between $t_{b1}$ and $t_{b2}$ to the peak-to-peak interval of the PPG waveform |

**Table 5.** Eighteen Demographic Time-domain Features

| Feature | Definition |
|---|---|
| 58. Height/$\Delta$T | It is known as stiffness index |
| 59. Weight/$\Delta$T | The ratio of weight to $\Delta$T |
| 60. BMI/$\Delta$T | The ratio of BMI to $\Delta$T |
| 61. Height/$t_1$ | The ratio of height to the systolic peak time |
| 62. Weight/$t_1$ | The ratio of weight to the systolic peak time |
| 63. BMI/$t_1$ | The ratio of BMI to the systolic peak time |
| 64. Height/$t_2$ | The ratio of height to the notch time |
| 65. Weight/$t_2$ | The ratio of weight to the notch time |
| 66. BMI/$t_2$ | The ratio of BMI to the notch time |
| 67. Height/$t_3$ | The ratio of height to the diastolic peak time |
| 68. Weight/$t_3$ | The ratio of weight to the diastolic peak time |
| 69. BMI/$t_3$ | The ratio of BMI to the diastolic peak time |
| 70. Height/$t_{pi}$ | The ratio of height to the pulse interval |
| 71. Weight/$t_{pi}$ | The ratio of weight to the pulse interval |
| 72. BMI/$t_{pi}$ | The ratio of BMI to the pulse interval |
| 73. Height/$t_{pp}$ | The ratio of height to the peak-to-peak interval |
| 74. Weight/$t_{pp}$ | The ratio of weight to the peak-to-peak interval |
| 75. BMI/$t_{pp}$ | The ratio of BMI to the peak-to-peak interval |

**Table 6.** Sixteen Frequency Domain Features

| Feature | Definition |
|---|---|
| 76. peak-1 | The amplitude of the first peak from the Fast Fourier Transform of the PPG signal |
| 77. peak-2 | The amplitude of the second peak from the Fast Fourier Transform of the PPG signal |
| 78. peak-3 | The amplitude of the third peak from the Fast Fourier Transform of the PPG signal |
| 79. Freq-1 | The frequency at which the first peak from the Fast Fourier Transform of the PPG signal occurred |
| 80. Freq-2 | The frequency at which the second peak from the Fast Fourier Transform of the PPG signal occurred |
| 81. Freq-3 | The frequency at which the third peak from the Fast Fourier Transform of the PPG signal occurred |
| 82. A0-2 | Area under the curve from 0 Hz to 2 Hz for the Fast Fourier Transform of the PPG signal |
| 83. A2-5 | Area under the curve from 2 Hz to 5 Hz for the Fast Fourier Transform of the PPG signal |
| 84. A0-2/A2-5 | The ratio of the area under the curve from 0 Hz to 2 Hz to the area under the curve from 2 Hz to 5 Hz |
| 85. pPeak-1/peak-2 | The ratio of the first peak to the second peak from the Fast Fourier Transform of the PPG signal |
| 86.Peak-1/peak-3 | The ratio of the first peak to the third peak from the Fast Fourier Transform of the PPG signal |
| 87. Freq-1/Ffreq-2 | The ratio of the frequency at first peak to the frequency at second peak from the Fast Fourier Transform of the PPG signal |
| 88. Freq-1/Ffreq-3 | The ratio of the frequency at first peak to the frequency at third peak from the Fast Fourier Transform of the PPG signal |
| 89. Maximum Frequency | The value of highest frequency in the signal spectrum. $$f_{max}$$ |
| 90. Magnitude at Fmax | Signal magnitude at highest Frequency. $$X(f_{max})$$ |
| 91. Ratio of signal energy | Ratio of signal energy between $(f_{max} \pm \Delta f)$ and the whole spectrum. $$X(f_{max} \pm \Delta f) / \sum_{i=0}^{N-1} X_i(f)$$ |

**Table 7.** Ten Statistical Features

| Feature | Definition | Equation |
|---|---|---|
| 92. Mean | Sum of all data divided by the number of entries | $$\bar{x} = \frac{\sum x}{n}$$ |
| 93. Median | Value that is in the middle of ordered set of data | Odd numbers of entries: Median=middle data entry. Even numbers of entries: Median=adding the two numbers in the middle and dividing the result by two. |
| 94. Standard Deviation | Measure variability and consistency of the sample. | $$s = \sqrt{\frac{\sum x - \bar{x}}{n-1}}$$ |
| 95. Percentile | The data value at which the percent of the value in the data set are less than or equal to this value. | $25^{th} = (\frac{25}{100})n$ <br> $75^{th} = (\frac{75}{100})n$ |
| 96. Mean Absolute Deviation | Average distance between the mean and each data value. | $$MAD = \frac{\sum_{i=1}^{n} |x_i - \bar{x}|}{n}$$ |
| 97. Inter Quartile Range (IQR) | The measure of the middle 50% of a data. | $IQR = Q_3 - Q_1$ |
| | | Q$_3$: third quartile, Q$_1$: first quartile, Quartile: dividing the data set into four equal portions. |
| 98. Skewness | The measure of the lack of symmetry from the mean of the dataset. | g1 = $\frac{\sum_{i=1}^{N}(Y_i - Y)^3/N}{S^3}$<br>Y: mean, s: the standard deviation, N: number of the data. |
| 99. Kurtosis | The pointedness of a peak in distribution curve, in other words it is the measures of sharpness of the peak of distribution curve. | K = $\frac{\sum_{i=1}^{N}(Y_i - Y)^4/N}{S^4} - 3$<br>Y: mean, s: the standard deviation, N: number of the data. |
| 100. Shannon's Entropy | Entropy measures the degree of randomness in a set of data, higher entropy indicates a greater randomness, and lower entropy indicates a lower randomness. | H(x)= -$\sum_{i=0}^{N-1} p_i \log_2 p_i$ |
| 101. Spectral Entropy | The normalized Shannon's entropy that is applied to the power spectrum density of the signal. | SEN = $\frac{-\sum_{i=0}^{N-1} p_k \log_2 p_k}{\log N}$<br>$p_k$: the spectral power of the normalized frequency, N: the number of frequencies in binary |

**Table 8.** Six Demographic Features

| 102. Height | 103. Weight | 104. Gender | 105. Age | 106. BMI | 107. Heart rate |
|---|---|---|---|---|---|

*2.4 Feature Selection*

Feature selection or reduction is important to reduce the risk of over-fitting the algorithms. In this work, three feature selection methods: correlation-based feature selection (CFS), ReliefF features selection [50], and features for classification using minimum redundancy maximum relevance (fscmrmr) algorithm. ReliefF is a feature selection algorithm, which randomly selects instances and adjusts the weights of the respective element depending on the nearest neighbor [51].

Correlation is a test used to evaluate whether or not a feature is highly correlated with the class or not highly correlated with any of the other features [52, 53]. On the other hand, the fscmrmr algorithm finds an optimal set of features that are mutually and as dissimilar as possible, and can effectively represent the response variable. The algorithm minimizes a feature set's inconsistency and maximizes the relevance of a feature set to the answer variable [54]. MATLAB built-in functions were used for CFS, ReliefF and fscmrmr feature selection algorithm [55].

In **Table 9**, the features selected by the feature reduction algorithm are listed. The features listed are those that produced the best results.

**Table 9:** Features chosen by the feature selection algorithms

| Feature Selection Algorithms Used | Systolic Blood Pressure | Diastolic Blood Pressure |
|---|---|---|
| RELIEFF | 105. Age,<br>106. Heart Rate,<br>103. Weight,<br>102. Height,<br>107. BMI,<br>83. A2-5, | 105. Age,<br>106. Heart Rate,<br>103. Weight,<br>102. Height,<br>107. BMI,<br>69. BMI/t$_3$, |

| | 63. BMI/$t_1$,<br>71. Weight/$t_{pi}$,<br>74. Weight/$t_{pp}$,<br>62. Weight/$t_1$,<br>75. BMI/$t_{pp}$ | 71.Weight/$t_{pi}$,<br>6. $t_3$,<br>72. BMI/$t_{pi}$,<br>82. A0-2 |
|---|---|---|
| FSCMRMR | 105. Age,<br>97. Inter Quartile Range,<br>45. $t_{a2}$,<br>64. Height/$t_2$,<br>13.Alternative Augmentation Index,<br>98. Skewness,<br>101. Spectral Entropy,<br>87. Freq-1/Freq-2,<br>23. x/($t_{pi}$-$t_1$),<br>32. Width(50%)/t1,<br>36. Width(50%)/tpi,<br>99. Kurtosis,<br>30. Width(25%)/$\Delta$T | 103. Weight,<br>22. $t_3$/y,<br>106. Heart Rate,<br>40. Width(75%)/$\Delta$T,<br>77. peak-2,<br>100. Shannon's Entropy,<br>96. Mean Absolute Deviation,<br>90. Magnitude at Fmax,<br>38. Width(75%)/$t_2$,<br>58. Height/$\Delta$T,<br>101. Spectral Entropy,<br>31. Width(25%)/tpi,<br>105. Age |
| CFS | 69. BMI/$t_3$,<br>71. Weight/$t_{pi}$,<br>74. Weight/$t_{pp}$,<br>49. $t_{b2}$,<br>59. Weight/$\Delta$T,<br>51. $b_1$/$a_1$,<br>46. $b_1$,<br>47$t_{b1}$,<br>62. Weight/$t_1$,<br>52. $t_{a1}$/$t_{pp}$,<br>66. BMI/$t_2$,<br>67. Height/$t_3$,<br>100(Shannon's Entropy),<br>48. $b_2$,<br>75. BMI/$t_{pp}$ | 69. BMI/$t_3$,<br>71. Weight/$t_{pi}$,<br>74. Weight/$t_{pp}$,<br>49. $t_{b2}$,<br>59. Weight/$\Delta$T,<br>51. $b_1$/$a_1$,<br>46. $b_1$,<br>47. $t_{b1}$,<br>62. Weight/$t_1$,<br>52. $t_{a1}$/$t_{pp}$,<br>66. BMI/$t_2$,<br>67. Height/$t_3$,<br>100. Shannon's Entropy,<br>48. $b_2$,<br>75. BMI/$t_{pp}$ |

*2.5 Machine Learning (ML) Algorithms*

After the features were extracted, feature matrix were trained with machine learning algorithms. Regression Learner App of MATLAB 2019b was used to estimate the BP. Five different algorithms (Linear Regression, Regression Trees, Support Vector Regression (SVR), Gaussian Process Regression (GPR), and Ensemble Trees) with their variations to a total of 19 algorithms were trained using 10-fold cross validation. Out of all these algorithms, two best performing algorithms, Gaussian Process Regression and Ensemble Trees were tested.

Gaussian Process Regression: GPR is a non-parametric Bayesian regression approach [56], which has benefits of operating well on small datasets and being able to provide measures of uncertainty on the predictions. Unlike many common supervised machine-learning algorithms that learn the exact values in a function for each parameter, the Bayesian approach infers a distribution of probability over all possible values.

Ensemble Trees: An ensemble tree is a predictive model consisting of a weighted combination of multiple regression trees [57]. The core idea behind the ensemble model is to pull together a set of weak learners to create a strong learner.

*2.6 Hyper-parameters Optimization of the Best Performing Algorithm*

The machine learning algorithms used were initially trained with default parameters. The performance of these algorithms can, however, be improved by optimizing their hyper-parameters. Hyper-parameters optimization was carried out on the algorithms using MATLAB 2019b Regression Learner App [58].

*2.7 Evaluation Criteria*

To evaluate the performance of the ML algorithms for estimating BP, four criteria were used. Here, Xp is the predicted data while the ground truth data is X and n is the number of samples:

Mean Absolute Error (MAE): Absolute Error is the amount of predicted error. The Mean Absolute Error is the mean of all absolute errors.

$$\text{MAE} = \frac{1}{n}\sum_n |X_p - X| \qquad (2)$$

Mean Squared Error (MSE): MSE calculates the squares sum of the errors. MSE is a risk function, which corresponds to the expected value of the squared error loss. MSE contains both the estimator's variance and its bias.

$$\text{MSE} = \frac{\sum |X_p - X|^2}{n} \qquad (3)$$

Root Mean Squared Error (RMSE): RMSE is the standard deviation of the residuals (prediction error). Residuals are a measure of how far away the data points are from the regression line; RMSE is a measure of how these residuals are spread out.

$$\text{RMSE} = \sqrt{\frac{\sum |X_p - X|^2}{n}} = \sqrt{MSE} \qquad (4)$$

Correlation Co-efficient (R): it is a statistical technique, which measures how closely related are two variables (predictors and the predictions). It also tells us how close the predictions are to the trendline.

$$\text{R} = \sqrt{1 - \frac{MSE(Model)}{MSE(Baseline)}}$$

$$\text{where, MSE (Baseline)} = \frac{\sum |X - mean(X)|^2}{n} \qquad (5)$$

When using the Regression Learner App in MATLAB, the above criteria are automatically calculated by MATLAB and these values were used to evaluate the performance of the algorithms. Among these criteria, RMSE was chosen as the main criterion.

**3. Results and Discussion**

This section summarizes the performance of the machine-learning algorithm used in the study. As stated earlier, 19 different machine-learning algorithms were trained and validated. It is observed from **Table 9** that the features of **Table 5** have significant contribution along with demographic features in estimation. Out of the 19 algorithms, GPR and Ensemble Trees outperformed for all cases in the estimation of both Systolic Blood Pressure and Diastolic Blood Pressure.

**Table 10.** Evaluation of the best performing algorithm for SBP and DBP

| Selection Criteria | Performance Criteria | Systolic Blood Pressure | | Diastolic Blood Pressure | |
|---|---|---|---|---|---|
| | | GPR | Ensemble Trees | GPR | Ensemble Trees |
| *Features from Literature* | *MAE* | 12.27 | 12.68 | 8.31 | 8.82 |
| | *MSE* | 240.25 | 246.74 | 96.90 | 109.92 |
| | *RMSE* | 15.50 | 15.70 | 9.84 | 10.45 |
| | *R* | 0.71 | 0.71 | 0.62 | 0.54 |
| *All Features (newly designed and from literature)* | *MAE* | 12.06 | 12.95 | 7.70 | 8.31 |
| | *MSE* | 272.32 | 316.71 | 97.31 | 110.87 |
| | *RMSE* | 16.50 | 17.80 | 9.86 | 10.53 |
| | *R* | 0.70 | 0.59 | 0.63 | 0.57 |
| *ReliefF* | *MAE* | **10.08** | 12.57 | 7.87 | 8.93 |
| | *MSE* | **219.08** | 258.16 | 96.70 | 119.32 |
| | *RMSE* | **14.80** | 16.06 | 9.83 | 10.92 |
| | *R* | **0.74** | 0.69 | 0.62 | 0.49 |
| *FSCMRMR* | *MAE* | 13.92 | 15.10 | 8.84 | 9.66 |
| | *MSE* | 302.75 | 349.06 | 112.27 | 128.43 |
| | *RMSE* | 17.39 | 18.68 | 10.59 | 11.33 |
| | *R* | 0.62 | 0.55 | 0.53 | 0.42 |
| *CFS* | *MAE* | 11.91 | 13.06 | **7.64** | 8.27 |
| | *MSE* | 257.77 | 325.29 | **83.95** | 103.70 |
| | *RMSE* | 16.05 | 18.03 | **9.16** | 10.18 |
| | *R* | 0.69 | 0.65 | **0.68** | 0.58 |

In **Table 10**, it can be noticed that ReliefF feature selection algorithm produced the best result when combined with GPR. Feature selected using ReliefF and GPR combination performed the best estimating SBP while CFS and GPR performed best for DBP. Moreover, R scored 0.74 and 0.68 for SBP and DBP respectively, which means that there is a strong correlation with the predictors and the ground truth. However, these results could be further improved by tuning the hyper-parameters. Bayesian Optimization was used, which is efficient and effective and operates by constructing a probabilistic model of the objective function, called the surrogate function, which is then optimally scanned with the acquisition function before the candidate samples are selected for evaluation of the real objective function. As shown in **Figure 13**, 30 iterations of the model were trained during optimization. Each time it iterates, it tunes the hyper-parameters. If the result gives an MSE, lower than the lowest MSE recorded, then that MSE is taken as the lowest. If there is no over-fitting, the lowest MSE should be reported at the end of the iterations.

**Table 11** summarizes the performances of the algorithms after optimization. It is clear that the ReliefF feature selection algorithm with GPR outperforms the other algorithms. After optimization, the combination produced a remarkable improvement in R score for SBP and DBP estimation (0.95/0.96).

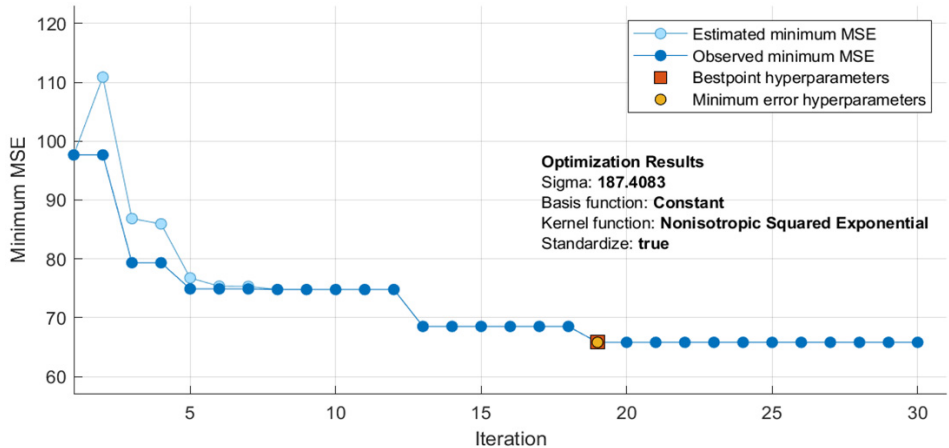

**Figure 13**: Optimization of GPR model during training.

**Table 11.** Evaluation of the outperforming algorithms for estimating SBP and DBP after Optimization.

| Selection Criteria | Performance Criteria | Systolic Blood Pressure | | Diastolic Blood Pressure | |
|---|---|---|---|---|---|
| | | Optimized GPR | Optimized Ensemble Trees | Optimized GPR | Optimized Ensemble Trees |
| *Features from Literature* | *MAE* | 6.79 | 12.43 | 4.49 | 8.17 |
| | *MSE* | 180.99 | 231.15 | 70.06 | 104.45 |
| | *RMSE* | 13.45 | 15.20 | 8.37 | 10.27 |
| | *R* | 0.79 | 0.73 | 0.74 | 0.57 |
| *All Features (newly designed and from literature)* | *MAE* | 3.30 | 10.886 | 2.81 | 7.96 |
| | *MSE* | 72.95 | 264.24 | 30.70 | 111.97 |
| | *RMSE* | 8.54 | 16.25 | 5.54 | 10.58 |
| | *R* | 0.92 | 0.67 | 0.90 | 0.56 |
| *ReliefF* | *MAE* | **3.02** | 11.32 | **1.74** | 5.99 |
| | *MSE* | **45.49** | 284.69 | **12.89** | 62.04 |
| | *RMSE* | **6.74** | 16.84 | **3.59** | 7.88 |
| | *R* | **0.95** | 0.65 | **0.96** | 0.78 |
| *FSCMRMR* | *MAE* | 6.11 | 14.65 | 6.80 | 8.22 |
| | *MSE* | 108.96 | 321.63 | 77.26 | 110.84 |
| | *RMSE* | 10.44 | 17.93 | 8.78 | 10.53 |
| | *R* | 0.88 | 0.58 | 0.72 | 0.56 |
| *CFS* | *MAE* | 12.95 | 16.27 | 7.59 | 7.89 |
| | *MSE* | 361.96 | 448.25 | 108.43 | 106.72 |
| | *RMSE* | 19.02 | 21.17 | 10.41 | 10.33 |
| | *R* | 0.50 | 0.28 | 0.57 | 0.58 |

In general, due to different evaluation criteria, and different and inadequately defined datasets, it is difficult to compare similar works in this field. Some reported lowest errors using small selected subsets of public or private data, but others worked on large-scale data (Kachuee et al. [24] and Slapničar et al. [30]) which has greater errors. Looking at individual related works in **Table 12,** Kachuee et al. [24] proposed method employs physiological parameters, machine learning and signal processing algorithms using PTT approach and some time-domain PPG Features, where they showed promising result according to British Hypertension Society (BHS). Kim et al. [23] compared artificial neural network (ANN) with multiple regressions as a BP estimation method , but their study is limited to 20 subjects only and did not identify DBP. Cattivelli et al. [25] introduced an algorithm for estimating BP, but used a very small amount of data (34 recordings for 25 subjects). Zhang et al. [27] described the SVM and neural network approach using time-domain features which is used directly for the study of BP regression, and good results were obtained compared to previous work.

**Table 12:** Comparison with related work in relations to dataset, methodology and estimation error

| Author | Method Used | Number of subjects | Performance criteria | Systolic Blood Pressure | Diastolic Blood Pressure |
|---|---|---|---|---|---|
| Kachuee et al. [24] | SVM | MIMIC II (1000 subjects) | MAE *MSE* *RMSE* *R* | 12.38 - - - | 6.34 - - - |
| Kim et al. [23] | Multiple non-linear regression (MLP) | 180 recordings, 45 subjects | MAE *MSE* *RMSE* *R* | 5.67 - - - | - - - - |
| Kim et al. [23] | artificial neural network ANN | 180 recordings, 45 subjects | MAE *MSE* *RMSE* *R* | 4.53 - - - | - - - - |
| Cattivelli et al. [25] | Proprietary Algorithm | MIMIC database (34 recordings, 25 subjects) | MAE *MSE* *RMSE* *R* | - 70.05 - - | - 35.08 - - |
| Zhang et al. [27] | Support Vector Machine (SVM) | 7000 samples from 32 patients | MAE *MSE* *RMSE* *R* | 11.64 - - - | 7.62 - - - |
| Zhang et al. [27] | Neural Network (9 input neurons) | 7000 samples from 32 patients | MAE *MSE* *RMSE* *R* | 11.89 - - - | 8.83 - - - |
| Zadi et al. [59] | Autoregressive moving average (ARMA) models | 15 subjects | MAE *MSE* *RMSE* *R* | - - 6.49 - | - - 4.33 - |
| Slapničar et al. [30] | Deep learning (Spectro-temporal ResNet) | MIMIC III database (510 subjects) | MAE *MSE* *RMSE* *R* | 9.43 - - - | 6.88 - - - |

| Su et al. [28]* | Deep learning (Long short-term memory (LSTM)) | 84 subjects | MAE | - | - |
|---|---|---|---|---|---|
| | | | *MSE* | - | - |
| | | | *RMSE* | 3.73 | 2.43 |
| | | | *R* | - | - |
| This work | Gaussian Process Regression (GPR) | 222 recordings, 126 subjects | MAE | **3.02** | **1.74** |
| | | | *MSE* | **45.49** | **12.89** |
| | | | *RMSE* | **6.74** | **3.59** |
| | | | *R* | **0.95** | **0.96** |

\* *Deep learning algorithm on a small database*

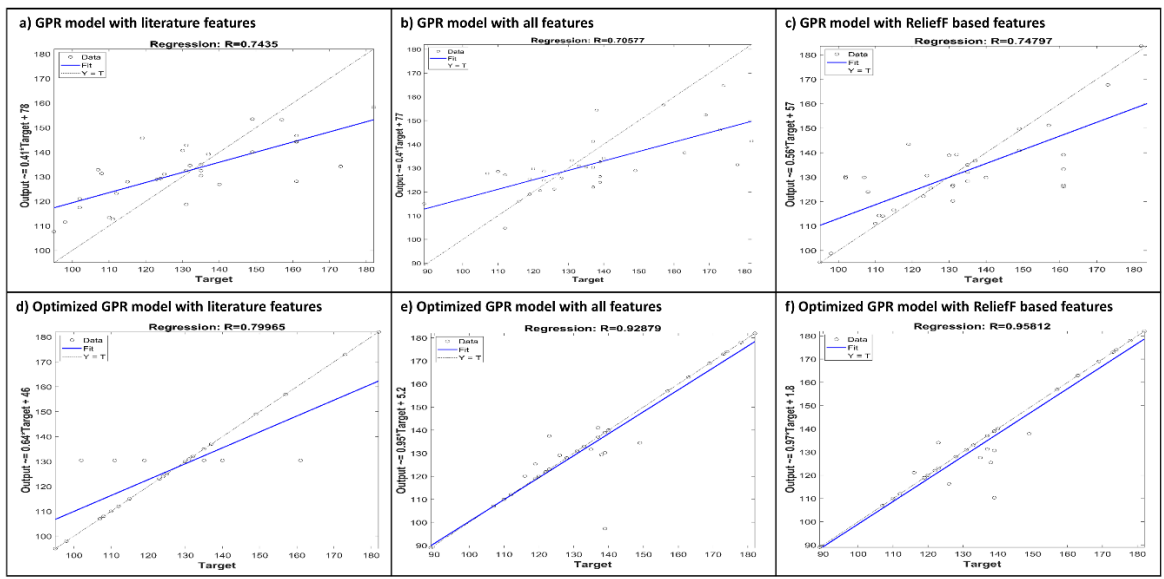

**Figure 14:** Comparison of the predicted output vs actual target for SBP estimation using different GPR: (a-c) models without optimization, (d-f) models with optimization.

In [59], Zadi et al. showed that the calculation of systolic and diastolic BP from PPG measurements using a viable method for continuous and non-invasive measurement of BP, however using a very small dataset (15 subjects only). Slapničar et al. [30] worked with a large dataset and using deep-learning Spectro-temporal ResNet algorithm has achieved a reasonable accuracy in estimation. Su et al. [28] used a conventional deep learning model for LSTM, but used the PTT approach as opposed to using only PPG on a small database. Finally, using time domain, frequency domain and statistical features to train an optimized feature reduced regression model, a very low error rate was achieved in this work. To the best of our knowledge, no work has extracted all these features and achieved such error rate using classical machine learning approach. In **Table 11**, comparative summary of recent works with this work is shown in respect of the evaluation parameters: MAE, MSE, RMSE and R.

It is also important to note that the standard for the evaluation of blood pressure measurement devices proposed by the Association for the Advancement of Medical Instrumentation (AAMI), the British Hypertension Society (BHS) and the International Organization for Standardization [60-63] is that a device is considered acceptable if the estimated blood pressure is less than 10 mmHg from the actual. The machine-learning algorithm proposed in the study has estimated with much higher precision and accuracy. According to **Table 13**, AAMI standard completely accepts the results of the GPR algorithm in DBP. However, the SD (standard deviation) of the model in the SBP evaluation is greater than the standard's maximum permissible range, but the mean is well in the acceptable range.

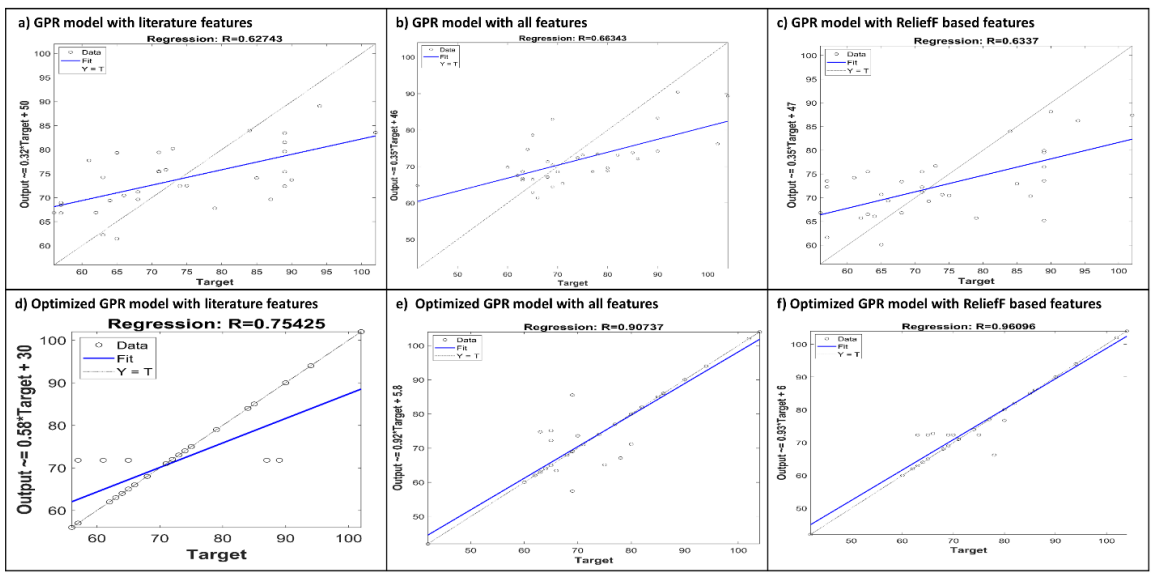

**Figure 15:** Comparison of the predicted output vs actual target for DBP estimation using different GPR: (a-c) models without optimization, (d-f) models with optimization.

**Table 13:** Comparison of this paper results with AAMI standard

|  |  | MEAN (mmHg) | SD (mmHg) | Subject |
|---|---|---|---|---|
| AAMI [62] | BP | ≤ 5 | ≤ 8 | ≥ 85 |
| This paper | SBP | 3.02 | 9.29 | 126 |
|  | DBP | 1.74 | 5.54 | 126 |

In addition, the accuracy of the proposed algorithm is tested from the point of view of the BHS grading criteria. Grades represent the cumulative percentage of readings falling within 5 mm Hg, 10 mm Hg, and 15 mm Hg of the mercury standard. The GPR algorithm findings are shown in **Table 14**, based on the BHS standard. The GPR model performance is consistent with the BHS standard grade B for both SBP and DBP estimation.

**Table 14:** Comparison of this paper results with BHS standard

|  |  | ≤ 5 mmHg | ≤ 10 mmHg | ≤ 15 mmHg |
|---|---|---|---|---|
| BHS [63] | grade A | 60% | 85% | 95% |
|  | grade B | 50% | 75% | 90% |
|  | grade C | 40% | 65% | 85% |
| This paper | SBP | 69% | 76% | 92% |
|  | DBP | 77% | 85% | 92% |

## 4. Conclusions

In this study, the authors have proposed and implemented a method for estimating Systolic and Diastolic blood pressure with the help of PPG signal features and machine learning algorithm. This successfully demonstrates how PPG signal can be used to accurately estimate the BP of patients non-invasively without using cuff-based pressure measurement. The entire pre-processing method of the PPG fingertip signals to extract the features, feature reduction and training of the algorithms were discussed. The raw signals were treated in different techniques and the resulting waveform has high signal-to-noise ratio and is free from baseline wandering. The system used time-domain, frequency-domain and statistical features and along with demographic data adding up to 107 features, to extract meaningful data. Models for SBP and DBP were trained separately as they often had different key features. 19 different machine-learning algorithms were trained for both SBP and DBP, out of which GPR and Ensemble Trees were the most promising. To reduce computational complexity, various feature selection methods were used. It was found that a combination of ReliefF feature selection and GPR machine learning algorithm produced the best result. However, hyper-parameter optimization was then used to improve the models further. The resulting models achieved a noteworthy R score of 0.95 and 0.96 for SBP and DBP respectively. DBP estimator fulfills the requirement of the AAMI standard while SBP estimator are following mean requirement but fall short than the standard deviation requirement by a small amount. SBP and DBP both fulfills grade B criteria according the BHS standard. In the future work, deep learning algorithms can be utilized with a larger dataset to produce better prediction model which can fulfill the A grade requirement of the BHS standard. The trained model can be used in developing commercial light computation-based prototypes that can accurately estimate the BP. Such a system can help in continuously monitoring BP and avoid any critical health conditions due to sudden changes.

**Author Contributions:** Experiments were designed by MHC, MNIS and MEHC. Experiments were performed by MHC, MNIS and MEHC. Results analysis, and interpretation and paper drafting were done by all authors.

**Funding:** The publication of this article was funded by the Qatar National Library and Qatar National Research Fund (QNRF) with the grant (NPRP12S-0227-190164).

**Acknowledgments:** The authors would like to thank the Qatar National Research Fund (QNRF) for the grant (NPRP12S-0227-190164) to bear the research personnel cost, which made this work possible.

**Conflicts of Interest:** The authors declare no conflict of interest.